\documentclass[twocolumn,aps,floatfix,prd,preprintnumbers]{revtex4}

\usepackage{epsfig}
\usepackage{amsmath}
\usepackage{latexsym}
\usepackage[psamsfonts]{amssymb}
\usepackage{graphicx}
\usepackage{ulem}
\usepackage{longtable}
\usepackage{epstopdf}
\usepackage{bm}
\usepackage{color}

\newcommand{\be}{\begin{equation}}
\newcommand{\ee}{\end{equation}}
\newcommand{\bea}{\begin{eqnarray}}
\newcommand{\eea}{\end{eqnarray}}
\newcommand{\bi}{\begin{itemize}}
\newcommand{\ei}{\end{itemize}}

\newcommand\QQbar{Q\bar{Q}}
\newcommand\mQ{m_{Q}}

\begin{document}


\title{Bethe-Salpeter wave functions of $\eta_c(2S)$ and $\psi(2S)$ states from full lattice QCD}


\author{Kazuki Nochi${}^{1}$}\email{nochi@nucl.phys.tohoku.ac.jp}
\author{Taichi Kawanai${}^{2}$}\email{t.kawanai@fz-juelich.de}
\author{Shoichi Sasaki${}^{1}$}\email{ssasaki@nucl.phys.tohoku.ac.jp}

\affiliation{${}^{1}$Department of Physics, Tohoku University, Sendai 980-8578, Japan}
\affiliation{${}^{2}$J\"ulich Supercomputing Center, J\"ulich D-52425, Germany} 

\date{\today}
\begin{abstract}
We discuss the internal structure of radially excited charmonium mesons 
based on the equal-time and Coulomb gauge Bethe-Salpeter (BS) amplitudes, 
which are obtained in lattice QCD. 
Our simulations are performed with a relativistic heavy-quark action for the charm quark
on the (2+1)-flavor PACS-CS gauge configurations at the lightest pion mass, 
$M_{\pi}=156(7)$ MeV.
The variational method is applied to the study of the optimal charmonium operators 
for ground and first excited states of $S$-wave charmonia. 
We successfully calculate the BS wave functions of $\eta_c(2S)$ and $\psi(2S)$ states, 
as well as $\eta_c(1S)$ and $J/\psi$ states, and then estimate 
the root-mean-square radii of both the $1S$ and $2S$ charmonium states. We also examine 
whether a series of the BS wave functions from the ground state to excited states can be 
described by a single set of the spin-independent and spin-dependent interquark potentials 
with a unique quark mass.
It is found that the quark kinetic mass and both the central and the spin-spin charmonium potentials,
determined from the $2S$ wave functions, fairly agree with the ones from
the $1S$ wave functions. This strongly supports the validity of the potential description for the charmonium system---at least, below the open-charm threshold.
\end{abstract}


\maketitle

 
\section{Introduction}
The constituent quark description of the heavy quarkonium
systems has been successful in aiding the qualitative understanding of 
properties of the charmonium and bottomonium states, especially 
below the thresholds for decays to mesons with 
open heavy flavor~\cite{{Eichten:1974af},{Godfrey:1985xj},{Barnes:2005pb}}.
The complicated dynamics of quarks and gluons in 
QCD could be described by the {\it interquark potential} 
within the framework of nonrelativistic 
quantum mechanics due to
the heavy degrees of freedom whose energy scale
is much higher than the QCD 
scale ($\Lambda_{\rm QCD}$).
Such nonrelativistic potential (NRp) models share a common feature
on the interquark potential that incorporates two types of underlying physics: 
the Coulombic potential, which dominates in short range 
in accordance with the asymptotic freedom of QCD, and a long-range potential 
responsible for the quark confinement~\cite{{Eichten:1974af},{Godfrey:1985xj},{Barnes:2005pb}}. 

For the heavy quarkonium systems,
the Cornell potential is often adopted, and its functional form is 
given by
\be
V(r)=-\frac{A}{r}+\sigma r + V_0\;,
\ee
where $A$ is the Coulombic coefficient, $\sigma$ denotes
the string tension, and $V_0$ is the constant term associated with
a self-energy contribution of the color sources~\cite{Eichten:1974af}.
If one-gluon exchange is responsible for the Coulombic term, 
the coefficient $A$ is associated with the strong coupling constant $\alpha_s$
as $A=\frac{4}{3}\alpha_s$. 

Although the Cornell potential was not directly derived from QCD, 
the functional form has been qualitatively justified by the static 
heavy quark potential obtained from Wilson loops in lattice QCD~\cite{Bali:2000gf}.
There are slight differences in terms of the Cornell parameters ($A$, $\sigma$)
between the phenomenological potential used in the NRp models 
and the Wilson-loop results. This is simply because the 
resulting {\it static potential} from Wilson loops is defined
in the infinitely heavy quark limit~\cite{Bali:2000gf}. 

The finite-mass correction should be somewhat 
taken into account for the Wilson-loop results.
Such corrections to the static potential are 
classified in powers of the inverse of both heavy quark mass $\mQ$ and quark momentum $\mQ v$
(relative quark velocity $v$) within the modern approach of effective field theory
called potential nonrelativistic QCD (pNRQCD)~\cite{Brambilla:2004jw}. 
In line with the framework of pNRQCD,  the ${\cal O}(1/\mQ)$ correction~\cite{Koma:2006si}
and the ${\cal O}(1/\mQ^2)$ spin-dependent corrections~\cite{Koma:2006fw} to the static 
potential have been computed in {\it quenched} QCD. 

The charm quark mass region is far outside the validity region for the $1/\mQ$ expansion~\cite{Bali:2000gf}.
Moreover, the convergence behavior seems to be questionable even at the bottom mass region.
Indeed, a spin-spin potential determined at ${\cal O}(1/\mQ^2)$ from lattice QCD exhibits 
a slightly ``attractive" interaction for the case of spin-triplet states~\cite{Koma:2006fw}.
It implies that a breakdown of the adiabatic approximation is not avoided even at the bottom sector, 
since the leading-order contribution of the spin-spin potential yields {\it wrong mass ordering among 
hyperfine multiplets}~\cite{Laschka:2012cf}.

An interesting idea to define a two-body potential from the equal-time Bethe-Salpeter (BS) amplitude
was proposed by Aoki, Hatsuda, and Ishii for studying the nuclear force 
in lattice QCD~\cite{{Ishii:2006ec},{Aoki:2009ji}}. 
Subsequently, Ikeda and Iida applied the same idea to the quarkonium system 
in order to compute the interquark potential without the 
adiabatic approximation~\cite{{Ikeda:2010nj},{Ikeda:2011bs}}. 
These preceding studies led us to propose a novel approach, where
both the quark kinetic mass and the interquark potential 
are self-consistently determined within the BS amplitude method, in order to 
obtain {\it proper interquark potential at finite quark mass} 
using lattice QCD~\cite{Kawanai:2011xb}.

We have elaborated the new approach to determine 
reliable interquark potential from lattice QCD 
in our previous works~\cite{{Kawanai:2011xb},{Kawanai:2011jt},{Kawanai:2013aca},{Kawanai:2015tga}}.
In this work, we will discuss the internal structure of the radially excited charmonium 
mesons as a further application of the BS amplitude method. 

It is frequently asked whether a universal interquark potential and a unique quark mass
can be simultaneously defined in a series of the BS amplitudes from the ground state to 
excited states.
This question is aimed at the validity of the potential description for the heavy quarkonium systems
and also the reliability of the interquark potential determined from the BS amplitudes.
In an attempt to settle these points, we will verify the validity of the BS amplitude method
through a direct comparison of the charmonium potentials which are independently evaluated 
from the BS wave functions of either the ground or first radially excited charmonium
mesons in lattice QCD simulations.

This paper is organized as follows: In Sec.~II, we describe an implementation
of the variational method~\cite{{Michael:1985ne},{Luscher:1990ck}} in the calculation 
of the BS wave functions for both the ground and radially 
excited states. Subsequently, we give a brief review of two methods for
determination of the interquark potential including the quark kinetic mass 
from the resulting BS wave functions. 
Section~III gives the numerical details in calculating the BS wave functions
of both the ground ($1S$) and first radially excited ($2S$) states for the $S$-wave charmonia.
In Sec.~III, we also discuss the validity of the potential description for 
the charmonium system through a direct comparison between 
the interquark potentials that are independently determined by both
the $1S$ and $2S$ charmonium states.
Finally, we close with a brief summary and our conclusions in Sec. IV.

\section{Formalism}

\subsection{BS wave functions determined through the variational method}
\label{sec2:bs_method_plus_variational_method}

The equal-time BS amplitude for the $n$th meson is defined by
\be
\phi_n({\bm r})=\langle 0|{\cal O}_{\QQbar}({\bm r})|n\rangle
\ee
with ${\bm r}$-dependent quark-antiquark operator ${\cal O}_{\QQbar}({\bm r})$, 
where ${\bm r}$ is the relative coordinate between a quark ($Q$) and antiquark ($\bar{Q}$)
at a certain time slice.
Although ${\cal O}_{\QQbar}({\bm r})$ can be defined in a gauge-invariant way, 
we hereafter consider the {\it Coulomb gauge BS amplitude}.
In the Coulomb gauge, the operator is simply given by ${\cal O}_{\QQbar}({\bm r})=
\sum_{\bm x}\bar{Q}({\bm x})\Gamma Q({\bm x}+{\bm r})$, 
where $\Gamma$ represents the Dirac $\gamma$ metrics.
The ${\bm r}$-dependent amplitude, $\phi_n({\bm r})$, in the rest frame is called 
the {\it  BS wave function}. Therefore, the BS wave function for the $n$th meson in the rest frame 
can be determined from  
the ${\bm r}$-dependent two-point correlation function constructed with
a usual quark bilinear operator ${\cal O}_\alpha$ and the ${\bm r}$-dependent one:
\bea
C_\alpha({\bm r}, t)&=&\langle 0|
{\cal O}_{\QQbar}({\bm r},t){\cal O}^{\dagger}_\alpha(0)
| 0 \rangle \nonumber \\
&=&\sum_{n}\langle 0|
{\cal O}_{\QQbar}({\bm r})| n\rangle\langle n|{\cal O}^{\dagger}_{\alpha}
| 0 \rangle e^{-tM_n} \nonumber \\
&=& \sum_{n}\phi_n({\bm r})V^{\ast}_{n,\alpha}e^{-tM_n} \;,
\label{Eq:Master}
\eea
where $M_n$ is the rest mass of the $n$th meson, and the ${\bm r}$-dependent amplitude 
$\phi_n({\bm r})$ corresponds to its BS wave function. 
The spectral amplitude $V_{n, \alpha}$ defined by $V_{n, \alpha}=\langle 0|{\cal O}_\alpha|n\rangle$
is introduced in the third line of Eq.~(\ref{Eq:Master}).
The correlation function $C_\alpha({\bm r}, t)$ clearly contains a superposition of orthogonal states.
The ground-state contribution is indeed isolated from those of the excited states in the large-$t$ region.

Although we only focused on the BS wave function of the ground state 
in the previous works~\cite{{Kawanai:2011xb},{Kawanai:2011jt},{Kawanai:2013aca},{Kawanai:2015tga}}, 
we here intend to obtain the BS wave function of {\it the radially excited states}. 
For this purpose, we adopt the variational method~\cite{{Michael:1985ne},{Luscher:1990ck}}
to find an optimal meson operator ${\cal O}_n^{\rm opt}$, 
which solely couples to a specific ($n$th) state in Eq.~(\ref{Eq:Master}), 
since $\langle0| {\cal O}^{\rm opt}_n| m\rangle \propto \delta_{n, m}$. 

Starting with a set of basis meson operators ${\cal O}_{\alpha}$ ($\alpha=1,..., N$),
we consider an $N\times N$ correlation matrix 
\be
G_{\alpha \beta}(t, 0)=\langle 0|{\cal O}_\alpha(t){\cal O}^\dagger_\beta(0)|0\rangle,
\label{Eq:MatrixCorr}
\ee
whose spectral decomposition is given by
\be
G_{\alpha \beta}(t, 0)=\sum_{n}V_{n,\alpha}V^{\ast}_{n,\beta}
e^{-tM_n}
\ee
with the spectral amplitude $V_{n,\alpha}$. 
Next, let us solve the generalized eigenvalue problem,
\be
G_{\alpha \beta}(t)\omega_{n, \beta}=\lambda_n(t, t_0)G_{\alpha \beta}(t_0)\omega_{n, \beta},
\label{Eq:GenEigen}
\ee
to obtain the $n$th eigenvalue $\lambda_n(t, t_0)$, where $t_0$ is a reference time slice,
and its eigenvector is $\omega_{n, \beta}$. 
If only the $N$ lowest states are propagating in the region where
$t\ge t_0$, the $n$th eigenvalue $\lambda_n(t, t_0)$ is given by
a single exponential form, with the rest mass of the $n$th meson as
\be
\lambda_n(t, t_0)=e^{-(t-t_0)M_n},
\ee
which corresponds to the eigenvalue of the transfer matrix
between two time slices $t$ and $t_0$.
Details of how to practically compute the eigenvalues $\lambda_n(t, t_0)$ 
are described in Appendix B of Ref.~\cite{Sasaki:2006jn}.

Simultaneously, one can obtain its $N$-dimensional eigenvector $\omega_{n, \alpha}$, which 
should be orthogonal to the spectral weight 
$\sum_\alpha \omega_{n,\alpha} V^{\ast}_{m, \alpha}=e^{M_n t_0/2} 
\delta_{n, m}$~\cite{{Sasaki:2006jn},{Gockeler:1994rx}}. 
Therefore, the optimal operator can be constructed 
by an appropriate linear combination of the basis meson operators ${\cal O}_\alpha$ with the eigenvector
$\omega_{n, \alpha}$:
\be
{\cal O}^{\rm opt}_n=\sum_{\alpha=1}^{N}\omega^{\ast}_{n, \alpha} {\cal O}_\alpha \;.
\ee

The $N$ types of the meson operator with fixed quantum number are, for instance, given by the quark bilinear 
operators composed of spatially smeared quarks with $N$ different smearing radii.
Subsequently, one can obtain the BS wave function of the $n$th meson state by using
$N$ types of ${\bm r}$-dependent two-point correlation functions constructed with the
$n$th eigenvalue
$\lambda_n$ and eigenvector $\omega_{n, \alpha}$ of Eq.~(\ref{Eq:GenEigen}) as below
\be
\phi_n({\bm r})=e^{M_n(t-t_0/2)}\sum_{\alpha=1}^N \omega^{\ast}_{n, \alpha}C_{\alpha}({\bm r}, t) \;.
\label{Eq:BSamplitude}
\ee
A similar procedure has been recently applied for a study of the inner structure of glueball states~\cite{Liang:2014jta}.

\subsection{Quark kinetic mass and interquark potential from BS amplitudes}
\label{sec2:kinetic_mass_potential}

In the past several years, we have demonstrated that the interquark potential and the quark kinetic mass,
both of which are key ingredients within the potential description of heavy-heavy and
heavy-light mesons, are successfully determined from the BS wave 
functions of {\it the ground state} of $S$-wave charmonium and 
charm-strange mesons~\cite{{Kawanai:2011jt},{Kawanai:2013aca},{Kawanai:2015tga}}.
A natural question arises: if we simply apply our proposed method on not only 
the ground state, but also its excited states, what result comes out, especially 
from the BS wave function of {\it radially excited states}? 

In this paper, we thus focus on the radially excited states of the $S$-wave meson states.
The Dirac $\gamma$ matrices ($\Gamma$) that appear in both operators
${\cal O}_{\QQbar}({\bm r})$ and 
${\cal O}_{\alpha}$ are chosen to be $\gamma_5$ for the pseudoscalar (PS) meson ($J^P=0^-$),
and $\gamma_i$ for the vector (V) meson ($J^P=1^-$). 
Recall that the spatial symmetry group on a lattice is reduced to
the octahedral point group $O_h$.
To take this into account, the ${\bm r}$-dependent BS wave function $\phi_n({\bm r})$
calculated by Eq.~(\ref{Eq:BSamplitude}) is supposed to be projected
in the $A_1^{+}$ representation, $\phi_n({\bm r})\rightarrow \phi_n(A_1^+; r)$, 
for $S$-wave mesons. Details of the $A_1^+$ projection 
are described in Ref.~\cite{Kawanai:2013aca}. 
Hereafter, the $A_1^{+}$ projected BS wave functions of $nS$ states 
for the PS and V channels are denoted by 
$\phi_{\rm PS}^{nS}(r)$ and $\phi_{\rm V}^{nS}(r)$.

In our preceding studies,
the quark kinetic mass $\mQ$ has been read off
from the long-distance asymptotic value 
of the difference of ``quantum kinetic energies" 
(the second spatial derivative of the BS wave function normalized by the BS wave function) 
between the spin-singlet (PS) and spin-triplet (V) states 
in the hyperfine multiplet 
for the $1S$ states~\cite{{Kawanai:2011xb},{Kawanai:2011jt},{Kawanai:2013aca},{Kawanai:2015tga}}.
We here generalize this idea to $nS$ states. The quark kinetic mass  can be determined from 
a set of the $nS$ wave functions in the following way~\cite{Kawanai:2011xb}:
\be
\mQ(nS) = \lim_{r \rightarrow \infty}\frac{-1}{E_{{\rm hyp}}(nS)}\left\{
\frac{{\bm \nabla}^2\phi_{\rm V}^{nS}(r)}{\phi_{\rm V}^{nS}(r)}
-
\frac{{\bm \nabla}^2\phi_{\rm PS}^{nS}(r)}{\phi_{\rm PS}^{nS}(r)}
\right\}
\label{eq:kinetic_mass}
\ee
with the hyperfine splitting energy of the $nS$ states, 
$E_{\rm hyp}(nS)=M_{\rm V}(nS)-M_{\rm PS}(nS)$.
The derivative ${\bm \nabla}^2$ that appears in Eq.~(\ref{eq:kinetic_mass}) 
is defined by the discrete Laplacian on the lattice. 
As shown in Ref.~\cite{Kawanai:2013aca}, a suitable choice of the discrete Laplacian is 
defined in the discrete polar coordinates in order to reduce the discretization artifacts 
on the short-range behavior of the interquark potential.

The interquark potential for $S$-wave states can be decomposed into
the central (spin-independent) potential $V_{\rm C}(r)$ and the spin-spin potential $V_{\rm S}(r)$, 
which are defined by the BS wave functions of $nS$ states,
$\phi_{\rm PS}^{nS}(r)$ and $\phi_{\rm V}^{nS}(r)$,
as below:
\bea
V_{\rm C}^{nS}(r)&=&E_{\rm ave}(nS)\nonumber \\
&+&\frac{1}{\mQ(nS)}\left\{
\frac{3}{4}\frac{{\bm \nabla}^2\phi_{\rm V}^{nS}(r)}{\phi_{\rm V}^{nS}(r)}
+
\frac{1}{4}\frac{{\bm \nabla}^2\phi_{\rm PS}^{nS}(r)}{\phi_{\rm PS}^{nS}(r)}
\right\}\;\;\;\;
\label{Eq:Center}
\eea
and
\bea
V_{\rm S}^{nS}(r)&=&E_{\rm hyp}(nS)\nonumber \\
&+&\frac{1}{\mQ(nS)}\left\{
\frac{{\bm \nabla}^2\phi_{\rm V}^{nS}(r)}{\phi_{\rm V}^{nS}(r)}
-
\frac{{\bm \nabla}^2\phi_{\rm PS}^{nS}(r)}{\phi_{\rm PS}^{nS}(r)}
\right\},
\label{Eq:SpinSpin}
\eea
where $E_{\rm ave}(nS)=M_{\rm ave}(nS)-2\mQ(nS)$. 
The mass $M_{\rm ave}(nS)$ denotes the spin-averaged mass 
for the $nS$ states as $\frac{3}{4}M_{\rm V}(nS)+\frac{1}{4}M_{\rm PS}(nS)$.

\subsection{``Time-dependent" method for the interquark potential}

A basic idea of Eqs.~(\ref{Eq:Center}) and (\ref{Eq:SpinSpin}) follows the method
developed by the HAL QCD Collaboration to derive hadron-hadron interactions
from lattice QCD~\cite{{Ishii:2006ec},{Aoki:2009ji}}. 
The original method advocated by the HAL QCD Collaboration starts 
from the fact that the equal-time BS wave function satisfies the ``stationary" Schr\"odinger equation
with a nonlocal and energy-independent potential below the inelastic threshold~\cite{Aoki:2009ji}.
We simply apply this method to the quark-antiquark ($\QQbar$) system~\footnote[22]{
In Ref.~\cite{Caswell:1978mt}, for QED bound states such as positronium or muonium, 
the reduction of the Bethe-Salpeter equation in quantum field theory to 
an equivalent Schr\"odinger equation in quantum mechanics was discussed 
in a systematic perturbation series. 
The wave functions and the interaction kernel are, however, described in momentum 
space rather than in coordinate space.}. 
Strictly speaking, no explicit energy dependence of the nonlocal potential in a finite box was 
proved only for the case of the short-range interaction~\cite{Aoki:2009ji}. In this sense, 
the $\QQbar$ system, where confining quark interaction is long ranged, does not
guarantee the existence of an energy-independent nonlocal potential even below 
open heavy-flavor thresholds. 

Thus, assuming the existence of an energy-independent nonlocal potential in the $\QQbar$ system~\footnote{If it is not the case, the quark kinetic mass obtained from 
Eq.(\ref{eq:kinetic_mass}) and the interquark potentials determined from Eqs.~(\ref{Eq:Center}) 
and (\ref{Eq:SpinSpin}) may just have the dependence of the choice of either the ground- or excited-state.}, 
let us consider the following ``time-independent" Schr\"odinger equation for the BS wave function
$\phi_{\Gamma}$
in the nonrelativistic approximation:
\be
\left\{
E_{\Gamma}+\frac{{\bm \nabla}^2}{\mQ}
\right\}
\phi_{\Gamma}({\bm r})
=\int d{\bm r}^{\prime}U({\bm r}, {\bm r}^\prime)\phi_{\Gamma}({\bm r}^\prime),
\label{Eq:TindepSE}
\ee
where $E_{\Gamma}=M_{\Gamma}-2\mQ$. 
As discussed in Refs~\cite{{Kawanai:2011xb},{Kawanai:2011jt}}, for the $S$-wave meson states, the local potentials $V_{\rm PS}(r)$ and $V_{\rm V}(r)$ defined
at the leading order of the velocity expansion, 
$U({\bm r}, {\bm r}^\prime)
=\left\{V_{\rm PS(V)}(r)+{\cal O}(v^2)\right\}\delta^2({\bm r}-{\bm r}^\prime)$
with $v=|{\bm \nabla}/\mQ|$, are given by 
\be
V_{\rm PS(V)}(r)=E_{\rm PS(V)}+\frac{1}{\mQ}
\frac{{\bm \nabla}^2\phi_{\rm PS(V)}(r)}{\phi_{\rm PS(V)}(r)} \;.
\ee
The interquark potential $V_{\rm PS(V)}(r)$ can be written by $V_{\rm PS(V)}(r)=V_{\rm C}(r)+({\bm S}_{Q}\cdot{\bm S}_{\bar{Q}})V_{\rm S}(r)$,
where the spin operator ${\bm S}_{Q}\cdot{\bm S}_{\bar{Q}}$ may 
be replaced by an expectation value of $-3/4$ $(1/4)$ for the PS(V) state. 
Then, $V_{\rm C}(r)$ and $V_{\rm S}(r)$ are separately obtained, 
as shown in Eqs.~(\ref{Eq:Center}) and (\ref{Eq:SpinSpin}). 

In our preceding works~\cite{{Kawanai:2011xb},{Kawanai:2011jt},{Kawanai:2013aca},{Kawanai:2015tga}},
the central and spin-spin interquark potentials are successfully extracted from 
the $1S$ meson states by using this method. We then use resulting 
potentials and quark masses as purely theoretical inputs so as to solve
the nonrelativistic Schr\"odinger equation for calculating accessible energy levels of charmonium and charmed-strange mesons without unknown parameters. The resultant spectra below the $D\bar{D}$ and $DK$
thresholds excellently agree with well-established experimental data~\cite{Kawanai:2015tga}. 

Starting with no explicit energy dependence on the nonlocal potential, 
the HAL QCD Collaboration has proposed an alternative method 
to derive the hadron-hadron interactions
by a so-called ``time-dependent" Schr\"odinger-like equation~\cite{HALQCD:2012aa}, 
instead of Eq.~(\ref{Eq:TindepSE}).
Here, we also may apply the new method to our $\QQbar$ system of interest. 
For this purpose, let us introduce the following correlation function:
\be
R_{\Gamma}({\bm r}, t)=C_{\Gamma}({\bm r}, t)/(e^{-\mQ t})^2,
\ee
where $\mQ$ denotes the quark kinetic mass that should be
determined in advance, as described in Eq.~(\ref{eq:kinetic_mass}). 

Considering the time derivative, $\frac{\partial}{\partial t}R_{\Gamma}({\bm r}, t)$, with the help of the 
spectral decomposition 
of the original ${\bm r}$-dependent correlation function $C_{\Gamma}({\bm r}, t)$, we then arrive
at the time-dependent Schr\"odinger-like equation for the $\QQbar$ system~\cite{HALQCD:2012aa} 
as well:
%
%
\begin{widetext}
\be
\left\{
\frac{1}{4\mQ}\frac{\partial^2}{\partial t^2}
-\frac{\partial}{\partial t}
+\frac{{\bm \nabla}^2}{\mQ}
\right\}R_{\rm PS(V)}({\bm r},t)
=\int d{\bm r}^{\prime}U({\bm r}, {\bm r}^\prime)R_{\rm PS(V)}({\bm r}^\prime,t),
\label{Eq:TdepSE}
\ee
\end{widetext}
where the first term on the left-hand side is responsible for the fully relativistic treatment for the kinetic term. 
For the $S$-wave mesons, the $A_1^{+}$ projection is supposed to be applied 
to the correlation functions defined above as 
$R_{\rm PS(V)}({\bm r}, t)\rightarrow R_{\rm PS(V)}(A_1^+; r, t)$.
Starting from Eq.~(\ref{Eq:TdepSE}) with the same approximation on the nonlocal potential $U({\bm r}, {\bm r}^\prime)$, we thus obtain the alternative formula of $V_{\rm PS(V)}(r)$ as follows:
\bea
V_{\rm PS(V)}(r)&=&\frac{1}{\mQ}\frac{{\bm \nabla}^2 R_{\rm PS(V)}(r, t)}{R_{\rm PS(V)}(r, t)}
-\frac{(\partial/\partial t) R_{\rm PS(V)}(r, t)}{R_{\rm PS(V)}(r, t)}
\nonumber\\
&&
+\frac{1}{4\mQ}\frac{(\partial/\partial t)^2 R_{\rm PS(V)}(r, t)}{R_{\rm PS(V)}(r, t)}\;.
\eea
We hereafter will omit the second derivative term of $t$ in the analysis, treating it on the same footing as 
was done in Eq.~(\ref{Eq:TindepSE}). Details of this relativistic correction will be discussed in a separate publication~\cite{futurepaper}.

%
%
\begin{table*}[ht]
\caption{Parameters of $(2+1)$-flavor dynamical QCD gauge field configurations
generated by the PACS-CS Collaboration~\cite{Aoki:2008sm}.
The columns list the number of flavors, the lattice volume, the $\beta$ value,
hopping parameters for light and strange quarks, approximate lattice spacing~(lattice cutoff),
spatial physical volume, pion mass, and the number of configurations to be analyzed. }

\label{tab:ensembles_full}
\begin{ruledtabular}                                                                                                                        
 \begin{tabular}{ccccccccc}
  $N_f$ & $N_s^3\times N_t$   &$\beta$ & $\kappa_{ud}$ &$\kappa_{s}$ &$a$~[fm]~($a^{-1}$ [GeV])    
  & $N_sa$~[fm]  & $M_\pi$~[MeV] & No. of configrations \\[2pt] \hline
  $2+1$ &$32^3\times 64$  &1.9     & 0.13781 & 0.13640 & 0.0907(13)~($\approx$ 2.18)  & 2.90(4) & $\approx$156 & 198\\ 
 \end{tabular}
\end{ruledtabular} 
\end{table*}

The most important feature
of the time-dependent approach is
that ``single-state dominance" is not necessarily achieved 
in the given correlation function $R_\Gamma(r,t)$~\cite{HALQCD:2012aa}.
Therefore, it enables us to use the data of $R_\Gamma(r,t)$ {\it in the earlier time range},
within the condition that the inelastic contribution is negligible in the entire $t$ region analyzed.
This advantage may lead to small statistical and systematic uncertainties on the final result.

There is one caveat: the new method for the $\QQbar$ system highly assumes 
that a series of the BS wave functions 
from the ground state to excited states is generated by the same 
``nonlocal" potential. 
In this work, we will later verify the validity of this method in the $\QQbar$ system 
through a direct comparison of the interquark potentials defined in
Eqs.~(\ref{Eq:Center}) and (\ref{Eq:SpinSpin}), which are independently determined 
from the BS wave functions of both the $1S$ and $2S$ charmonium states.

\section{Numerical results}

The computation of the BS wave functions for the charmonium system is
carried out on a lattice, $N_s^3\times N_t = 32^3 \times 64$ using the (2+1)-flavor
PACS-CS gauge configurations, where the simulated pion mass is closest to 
the physical point as $m_{\pi}=156(7)$ MeV~\cite{Aoki:2008sm}.
Simulation parameters of PACS-CS gauge configurations are summarized in
Table~\ref{tab:ensembles_full}. Our results are analyzed on all 198 gauge
configurations, which are available through International Lattice Data
Grid and the Japan Lattice Data Grid~\footnote{International Lattice Data Grid/Japan Lattice Data Grid,
http://www.jldg.org.}. 

For the charm quark, we employ
the relativistic heavy quark (RHQ) action that removes the main discretization errors
induced by large charm quark mass. The RHQ action, which is a variant of 
the Fermilab approach~\cite{ElKhadra:1996mp}, 
is the anisotropic version of the ${\cal O}(a)$ improved Wilson action with 
five parameters $\kappa_c$, $\nu$, $r_s$, $c_B$, and $c_E$, called {\it RHQ parameters}
(for more details, see Refs.~\cite{{Aoki:2001ra},{Kayaba:2006cg}} ).

%
%
\begin{table}[hb]                                                                                       
  \caption{
    The hopping parameter $\kappa_c$ and the RHQ parameters ($\nu$, $r_s$, $c_B$, and  $c_E$)
    used for the charm quark. 
      \label{tab:RHQ_para_charmonium}
      }
      \begin{ruledtabular}                                                                              
      \begin{tabular}{ccccc} 
	$\kappa_c$ & $\nu$ & $r_s$  & $c_B$ & $c_E$  \\ \hline
	0.10819 & 1.2153 & 1.2131  & 2.0268 & 1.7911  
      \end{tabular}
     \end{ruledtabular}                                                                                
\end{table}

The parameters $r_s$, $c_B$, and $c_E$ in the RHQ action are determined by tadpole improved
one-loop perturbation theory~\cite{Kayaba:2006cg} with a reference of
the ${\cal O}(a)$ improvement coefficient, $c_{\rm SW}=1.715$ for light
quarks~\cite{Aoki:2008sm}. As for $\nu$, we use a nonperturbatively
determined value, which is tuned by reproducing the effective speed of light as unity
in the dispersion relation for the spin-averaged $1S$-charmonium state, since
the parameter $\nu$ is sensitive to the size of hyperfine splitting energy~\cite{Namekawa:2011wt}.
Our chosen RHQ parameters are summarized in Table~\ref{tab:RHQ_para_charmonium}.

When the quark propagator is computed, Dirichlet boundary conditions are imposed for the time direction 
at $t/a=0$ and 63 to eliminate unwanted contributions across time boundaries. 
Source location is set at two different time slices, $t_{\rm s}/a=6$ and 57, both of which are the same 
distance away from two boundaries, so as to avoid the temporal boundary effect. 
Averaging the results of calculations over multiple sources would help to reduce the statistical uncertainties.
Instead of changing the places of the boundaries and source locations, a temporal shift can be applied to the gauge configurations 
$\{U_{\mu}({\bm x},t )\}\rightarrow \{U_{\mu}({\bm x},t+t_{\rm shift})\}$ due to the temporal periodicity of the lattice.

We use gauge-covariant, approximately Gaussian-shaped smearing~\cite{{Gusken:1989qx},{Alexandrou:1992ti}} for constructing the spatially smeared operator 
${\cal O}_{\alpha}(x)=\bar{\cal Q}_{\alpha}(x)\Gamma {\cal Q}_{\alpha}(x)$ with
\be
{\cal Q}_{\alpha}({\bm x}, t)=\left(1+\frac{W_{\rm G}^2}{4N_{\rm G}}{{\bm D}^2}\right)^{N_{\rm G}} Q({\bm x}, t),
\ee
where ${\bm D}^2$ denotes the covariant lattice Laplacian, and $\alpha$ labels a set of 
two parameters as $\alpha=\{N_{\rm G}, W_{\rm G}\}$~\cite{Berruto:2005hg}. Here, $N_{\rm G}$ is the number of times the smearing kernel acts on the quark fields, while $W_{\rm G}$ is the width of the Gaussian that results 
in $N_{\rm G}\rightarrow \infty$.
We adopt four parameter sets: $\{N_{\rm G}, W_{\rm G}\}=\{10, 1.0\}$, $\{15, 2.0\}$, $\{20, 3.0\}$, and 
$\{30, 4.0\}$, so as to construct the $4 \times 4$ correlation matrix defined 
in Eq.~(\ref{Eq:MatrixCorr}) and also four types of
${\bm r}$-dependent two-point correlation function, defined in Eq.~(\ref{Eq:Master}).

  %
  %
  \begin{figure*}
  \centering
  \includegraphics*[width=.48\textwidth]{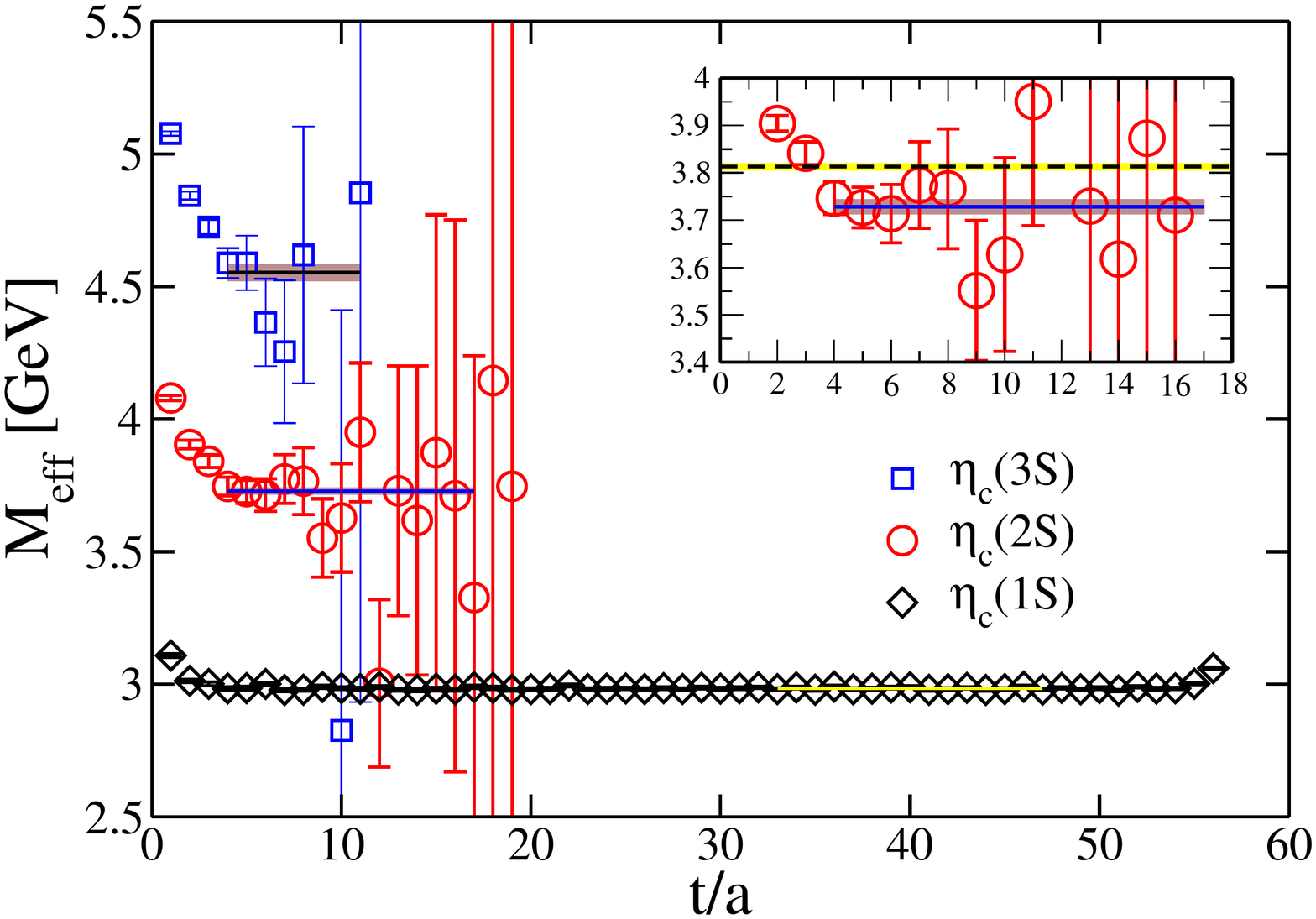} 
  \includegraphics*[width=.48\textwidth]{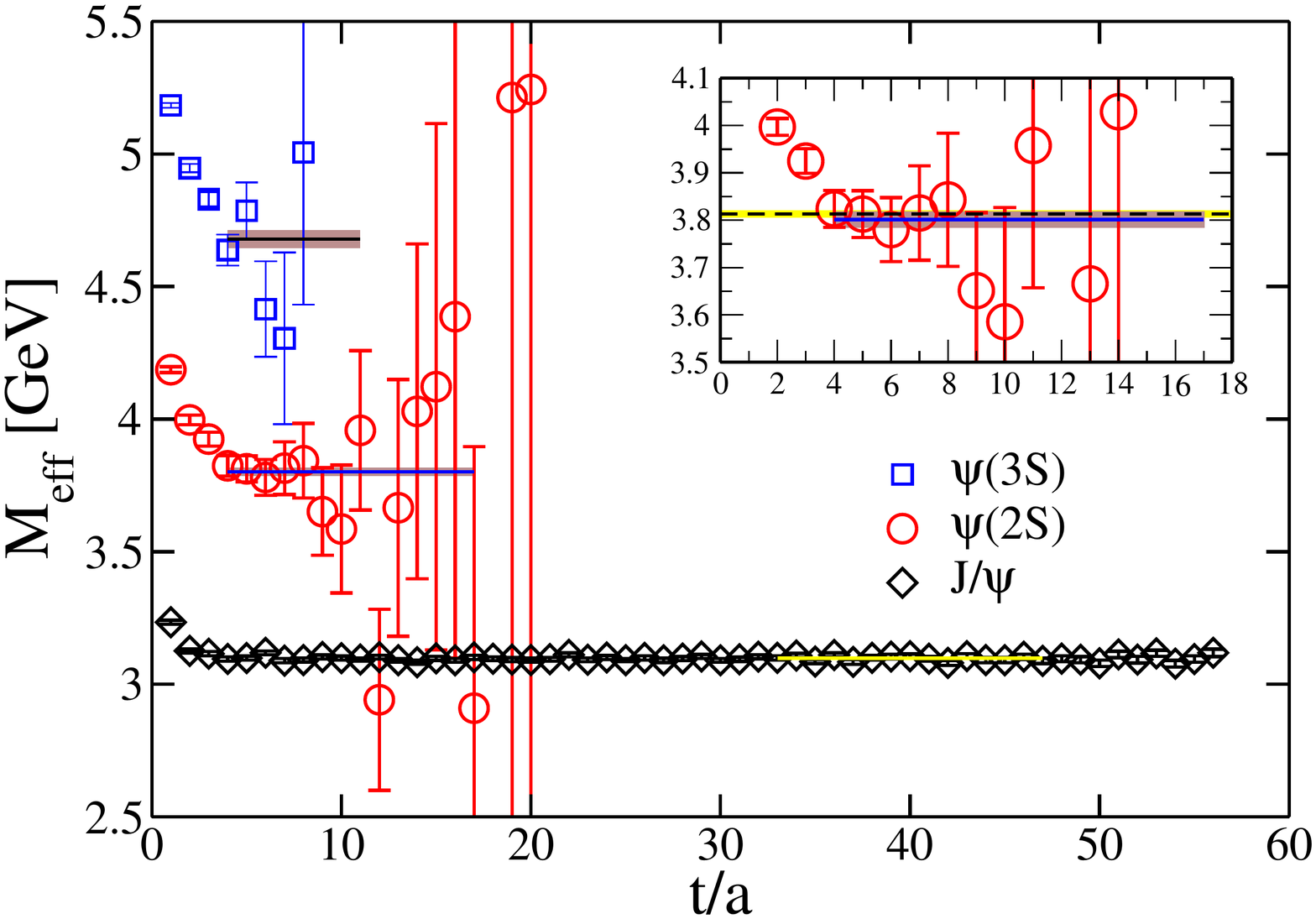} 
     \caption{Effective mass plots for the $\eta_c(nS)$ (left panel) and $\psi(nS)$ (right panel) states.
     Charmonium states are specified in the legend. Solid lines indicate fit results, and 
     shaded bands display the fitting ranges and 1 standard deviation. 
     The inset in each panel shows a magnified view of the effective mass of the 
     $2S$ state together with the $P$-wave $D\overline{D}$ threshold (dashed line).
     }
  \label{fig:EffectiveMass}
  \end{figure*}

For a single-parameter set, we compute 32 valence quark propagators per gauge configuration with 
eight different spatial centers of the Gaussian sources, which are located at the corners of a $16^3$ cube, 
on two different time slices $t_{\rm s}/a=6$ and 57, using two different 
temporal shifts $t_{\rm shift}/a=0$ and 32, so as to increase statistics.  
All 32 sets of usual and ${\bm r}$-dependent two-point correlation functions are 
folded together to create the single-correlation functions as a function of 
$t^\prime$ defined in the range $0\le t^\prime/a \le 57$. Hereafter, $t^\prime$ 
is simply denoted as $t$.

Let us first present the effective masses of the $S$-wave charmonium ($\eta_c$ and $\psi$) states
in the variational method. An effective mass is defined as 
\begin{equation}
M_{n, \Gamma}(t)=\log\frac{\lambda_{n, \Gamma}(t, t_0)}{\lambda_{n, \Gamma}(t+a, t_0)},
\end{equation}
where $\lambda_{n, \Gamma}(t, t_0)$ is the $n$th eigenvalue of the $4 \times 4$ correlation matrix
for $\Gamma={\rm PS}$ or V.
In this study, we choose the reference time slice as $t_0/a=3$, where the resulting mass
is less sensitive to variation of $t_0$.

%
%
\begin{table}[ht]                                                                                       
  \caption{
    Masses of $S$-wave charmonium states calculated from eigenvalues of the $4\times 4$ transfer matrix 
    up to $3S$ states. The fitting ranges and values of $\chi^2/\text{d.o.f.}$ are also included.
    For $1S$ and $2S$ charmonium states, the spin-averaged mass ($M_{\rm ave}$)
    and hyperfine splitting energy ($E_{\rm hyp}$) are also evaluated. Results are given in units of GeV.
    \label{tab:charmonium_mass}
      }
      \begin{ruledtabular}                                                                              
      \begin{tabular}{ccclc} 
       State & $J^{PC}$ & Fit range & Mass~[GeV] & $\chi^2/\text{d.o.f.}$\\ \hline
      $\eta_c(1S)$ & $0^{-+}$    & [33:47] & 2.9850(5) & 1.08\\
      $\eta_c(2S)$ & $0^{-+}$    & [4:17] & 3.729(15) & 0.80\\
      $\eta_c(3S)$ & $0^{-+}$   & [4:11] & 4.553(34) & 0.77\\
      $J/\psi$     & $1^{-+}$       &  [33:47] & 3.0986(14)  & 1.21 \\ 
      $\psi(2S)$ & $1^{-+}$       & [4:17] & 3.801(16) & 1.05\\ 
      $\psi(3S)$ & $1^{-+}$       & [4:11] & 4.679(34) & 1.53\\ 
      $M_{\text{ave}}(1S)$ & $\cdot\cdot\cdot$  & $\cdot\cdot\cdot$ & 3.0702(11)  & $\cdot\cdot\cdot$ \\ 
      $M_{\text{ave}}(2S)$ & $\cdot\cdot\cdot$  & $\cdot\cdot\cdot$ & 3.783(15)    & $\cdot\cdot\cdot$ \\ 
      $E_{\text{hyp}}(1S)$  & $\cdot\cdot\cdot$  & $\cdot\cdot\cdot$ & 0.1135(12)  & $\cdot\cdot\cdot$ \\
      $E_{\text{hyp}}(2S)$  & $\cdot\cdot\cdot$  & $\cdot\cdot\cdot$ & 0.0725(56)  & $\cdot\cdot\cdot$ \\
    \end{tabular}
      \end{ruledtabular}                                                                                
\end{table}

Figure~\ref{fig:EffectiveMass} shows the effective mass plots of 
the first three eigenvalues $\lambda_{1, \Gamma} > \lambda_{2, \Gamma} > \lambda_{3, \Gamma}$ 
for the PS and V channels. Here, we remark that $\lambda_{n, {\rm PS(V)}}$ is associated with the $nS$ state.
The variational method with the correlation matrix constructed in our chosen basis
successfully separates the first excited ($2S$) state and the second excited ($3S$) state 
from the ground ($1S$) state.

The horizontal solid lines represent each fit result with its 1 standard deviation obtained
by a covariant single exponential fit. In Table~\ref{tab:charmonium_mass}, 
we summarize the results of masses of the three lowest-lying $S$-wave charmonium states 
together with fit ranges used in the fits and values of $\chi^2$ per degrees of freedom (d.o.f.). 
 
The fit results for $3S$ states are rather sensitive to the choice of the fit range, 
since the signal of the $3S$ states dies out quickly; therefore, 
those values involved in Table~\ref{tab:charmonium_mass} are just listed for reference. 
The errors quoted in all of the results represent only the statistical errors given by the jackknife analysis.

For the $1S$ states, all results including $M_{\rm ave}$ and $E_{\rm hyp}$ 
obtained in the variational method are fully 
consistent with our previous study, where the charm quark propagators were computed 
by the wall source with the Coulomb gauge fixing~\cite{Kawanai:2015tga}.
It is worth recalling that the values of $\kappa_c$ and $\nu$ in the RHQ parameters are chosen
to reproduce both the experimental spin-averaged mass and hyperfine splitting energy 
of $1S$ charmonium states. This is the reason why our results of the $1S$ states 
are very close to the experimental values.

On the other hand, the masses of the $2S$ states correspond to the theoretical 
predictions from dynamical lattice QCD. 
We obtain results of $M_{\eta_c(2S)}=3.729(15)(21)$~GeV and $M_{\psi(2S)}=3.801(16)(31)$~GeV.
The first errors are statistical, and the second errors are systematic uncertainties due to
variations of $t_{\rm min}$ in the fit range [$t_{\rm min}/a:t_{\rm max}/a$].

Although those values are about 100~MeV higher 
than the experimental values of $M^{\rm exp}_{\eta_c(2S)}=3.639$~GeV 
and $M^{\rm exp}_{\psi(2S)}=3.686$~GeV, similarly higher values are reported in Ref.~\cite{Mohler:2011ke}.
In addition, the hyperfine splitting energy of the $2S$ states is $M_{\psi(2S)}-M_{\eta_c(2S)}=73(6)(1)$~MeV, 
of which the value is slightly larger than the experimental value of 47~MeV.
Needless to say, the higher-lying states might suffer much from the lattice artifacts---
finite size and lattice discretization effects---compared to their ground state~\footnote{
It is reminded that masses of $1P$-charmonium states ($\chi_{c0}$, $\chi_{c1}$ and $h_c$)
in the current set-up are fairly consistent with the experimental values, 
as studied in Refs.~\cite{{Kawanai:2011jt},{Kawanai:2015tga}}.}.

Our results for the $\eta_c(2S)$ and $\psi(2S)$ masses are near to and slightly above
the experimental value of the $D\bar{D}$ threshold energy ($\sim 3.730$~GeV). 
We, however, remark that since the $\eta_c$ and $\psi$ mesons have negative parity, 
the $P$-wave $D\bar{D}$ threshold energy, which is defined as the total energy of the noninteracting
$D\bar{D}$ state with the smallest nonzero momentum $|{\bm p}_{\rm min}|=2\pi/(La)$, 
is appropriate for comparison with the $\eta_c(2S)$ and $\psi(2S)$ masses~\cite{Sasaki:2003gi}.
In our calculation, the lowest open charm threshold is 3.813(8)~GeV, which is determined with 
the measured $D$-meson mass [$M_{D}=1.858(4)$~GeV]. Our result for 
the $\psi(2S)$ mass is slightly below but close to the $P$-wave $D\bar{D}$ threshold, while
the $\eta_c(2S)$ mass is well below the $P$-wave $D\bar{D}$ threshold~\footnote{
The $P$-wave $DD^*$ and $D^*\bar{D}^*$ threshold energies are determined as
$3.973(12)$~GeV and $4.133(17)$~GeV with the measured $D^*$ meson mass [$M_{D^*}=2.022(8)$~GeV]
in our calculation. These are much above our results of the $\eta_c(2S)$ and $\psi(2S)$ masses.}.
In this context, it would be important to know how much the $D\bar{D}$ mixing
effect has affected the spectroscopy of the $2S$ charmonium states.

Although more systematic study is thus necessary for the spectroscopy of the radially excited states,
it is beyond the scope of this paper. Rather, our main purpose is practically
to get the optimal charmonium operators
for both $1S$ and $2S$ states using the resulting eigenvectors $(\omega_{n, \Gamma})_{\alpha}$ 
of the transfer matrix in the variational method. As a result, the BS wave functions for $1S$ and $2S$ states
are obtained through Eq.~(\ref{Eq:BSamplitude}) separately.

\begin{figure}[ht]
  \centering
   \includegraphics*[width=.50\textwidth]{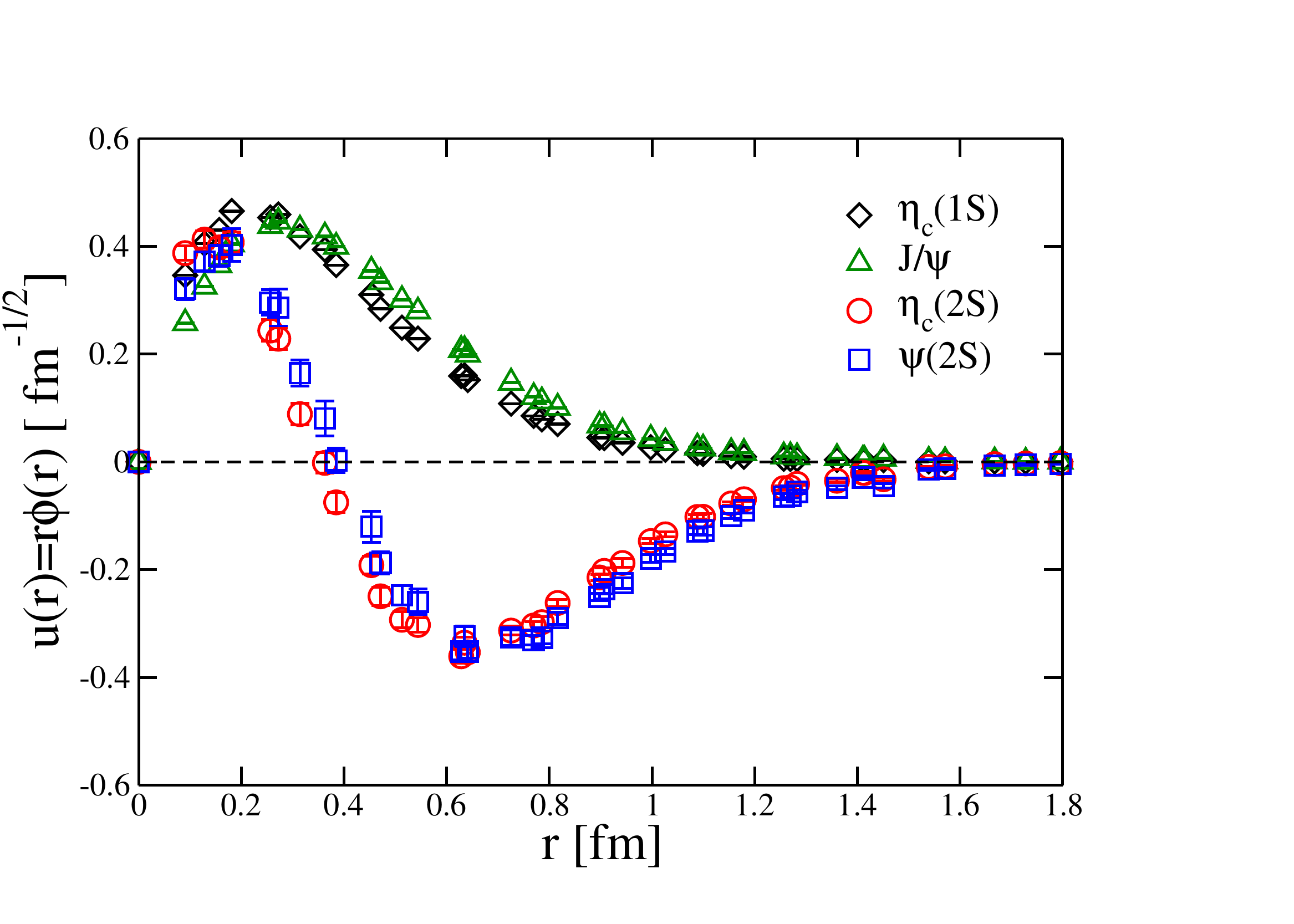} 
     \caption{The reduced BS wave functions $u(r)=r\phi(r)$ for the $\eta_c(1S)$ 
     (diamonds), $J/\psi$ (triangles), $\eta_c(2S)$ (circles), and
     $\psi(2S)$ (squares) states, shown as functions of the spatial distance $r$. 
     They are normalized as $\sum_{\bm r}|\phi({\bm r})|^2=1$. 
     The time average was performed in the range of $[t_{\rm min}/a:t_{\rm max}/a]=[24:33]$
     for the $1S$ states and [7:12] for the $2S$ states.
     }
  \label{fig:BSwavefunc}
  \end{figure}

In Fig.~\ref{fig:BSwavefunc}, we show the reduced wave functions 
$u_{n, \Gamma}(r)=r\phi_{n, \Gamma}({\bm r})$ 
of both $1S$ and $2S$ charmonium states for displaying the spatial distribution of the BS wave function.
The wave functions displayed in Fig.~\ref{fig:BSwavefunc} are normalized 
as $\sum_{\bm r}|\phi_{n, \Gamma}({\bm r})|^2=1$~\footnote{
For the $S$-wave states, we simply use a relation of 
$\sum_{\bm r}|\phi_{n, \Gamma}({\bm r})|^2=4\pi\int dr |u_{n, \Gamma}(r)|^2$ and 
then perform one-dimensional numerical integration in $r$ space by both the Simpson 1/3 
formula and the trapezoid formula in this study.}.
We plot data points taken along simpler ${\bm r}$ vectors, which are multiples of 
three directions---(1,0,0), (1,1,0), and (1,1,1)---
in order to avoid large discretization errors induced by the discrete 
Laplacian ${\bm \nabla}^2$~\cite{Kawanai:2013aca} in later discussion. 

Compared with the results of $1S$ states, the BS wave functions of 
both $\eta_c(2S)$ and $\psi(2S)$ states exhibit a specific nodal structure 
in the radial direction, as we expected.
Although at first glance the $2S$ wave functions are slightly extended in space
in comparison to the $1S$ wave functions, the spacial lattice extent $N_s a\approx 2.9$~fm is
likely to be large enough to study even the $2S$ charmonium system as well as the ground-state
charmonium states. 

The wave function provides information about a spatial size of the charmonium meson as
the root-mean-square (rms) radius $r_{\rm rms}$, which can be determined by
\be
r_{\rm rms}^2=\frac{\sum_{{\bm r}} r^2|\phi_{n, \Gamma}({\bm r})|^2}{\sum_{{\bm r}}|\phi_{n, \Gamma}({\bm r})|^2}=\frac{\int dr r^2|u_{n, \Gamma}(r)|^2}{\int dr |u_{n, \Gamma}(r)|^2}.
\ee
We then obtain the smaller rms radii for $1S$ states as
$(r_{\rm rms})_{\overline{1S}}\sim$ $0.38$~fm, while
$2S$ states yield comparatively larger values as $(r_{\rm rms})_{\overline{2S}}\sim$  $0.60$~fm. 
Another important aspect of the resulting $r_{\rm rms}$ is that both the $1S$ and $2S$ states
satisfy the relation $r_{\rm rms, PS} < r_{\rm rms, V}$. This simply indicates 
the repulsive nature of spin-spin interaction near the origin for the higher spin states.
All results of $r_{\rm rms}$ are summarized in Table~\ref{tab:charmonium_rms}.

  %
  %

We also verify the orthonormality condition between the resulting $1S$ and $2S$ wave functions through
the following overlap (OVL) coefficient:
\be
C_{\rm OVL}
=\frac{\sum_{{\bm r}} \phi^{1S}_{\Gamma}({\bm r})\phi^{2S}_{\Gamma}({\bm r})}{\sqrt{\sum_{{\bm r}}|\phi^{1S}_{\Gamma}({\bm r})|^2}\sqrt{\sum_{{\bm r}}|\phi^{2S}_{\Gamma}({\bm r})|^2}}.
\ee
We then obtain $|C_{\rm OVL}|=0.174(46)(3)$ 
for the PS channel and $|C_{\rm OVL}|=0.101(52)(6)$ 
for the V channel. The first error is statistical, and the second one is systematic uncertainty due to the choice of the numerical integral methods in $r$ space. Nonzero values of $|C_{\rm OVL}|$ in both the PS and V channels
suggest that the $2S$ wave function may receive non-negligible contamination of the $1S$ state from an unknown origin. 

There is, however, a hint from Fig.~\ref{fig:EffectiveMass}. 
Around $t/a=8$, the signal of the $2S$ states in the effective mass plot becomes noisy, 
and the isolation of the $2S$ states is statistically insignificant due to the large uncertainties.
Even if the eigenvector for the $2S$ states, $\omega_{2S, \alpha}$, is properly calculated in the variational method, contributions of the $2S$ state in the correlation $C_{\alpha}({\bm r}, t)$ are 
exponentially suppressed by its large mass $M_{2S}(>M_{1S})$ as a function of $t$. 
Therefore, if we include the data points of the $2S$ wave function determined at the larger $t$ 
during the averaging process over the time-slice range, the resulting $2S$ wave function may receive
a little component of the $1S$ wave function that is caused by incomplete orthogonal
factorization within numerical precision and its enhancement due to the relative suppression
of the $2S$-state contribution in the large-$t$ region.
Indeed, we observe that the overlap coefficient gets away from zero as the value of $t_{\rm min}$ increases in the time-averaged procedure. 

The small but nonzero value of $|C_{\rm OVL}|$ may cause serious systematic error in the early estimation 
of the rms radii for $2S$ states.
Taking into account such contaminations in the resulting $2S$ wave functions, 
we perform the Gram-Schmidt orthonormalization (GSO) so as to get an exactly 
orthogonal wave function to the $1S$ state as
\be
\tilde{\phi}^{2S}_{\Gamma}({\bm r}) = \frac{\phi^{2S}_{\Gamma}({\bm r})}
{\sqrt{\sum_{{\bm r}}|\phi^{2S}_{\Gamma}({\bm r})|^2}}
 - C_{\rm OVL} \times \frac{\phi^{1S}_{\Gamma}({\bm r})}{\sqrt{\sum_{{\bm r}}|\phi^{1S}_{\Gamma}({\bm r})|^2}}.
\ee
We then recalculate the rms radii of the $2S$ states with the above modified $2S$ wave function 
$\tilde{\phi}^{2S}_{\Gamma}({\bm r})$ for each channel.
We obtain slightly larger values as $(r_{\rm rms})_{\overline{2S}, {\rm GSO}}\sim$  $0.63$~fm in comparison to
the ones obtained from the original $2S$ wave functions.
The modified results of $r_{\rm rms}$ for $2S$ states are also included in Table~\ref{tab:charmonium_rms}.

In the following discussions, we do not use the modified $2S$ wave functions and keep the original ones
for our later analysis, since there is only a slight difference in their profile shapes, which mainly appears
at short distances, between the $2S$ wave functions obtained before and after the Gram-Schmidt orthonormalization.

%
%
\begin{table}[ht]                                                                                       
  \caption{
    Summary of the rms radii of $1S$ and $2S$ charmonium states, which are evaluated
    from the BS wave functions on the lattice.
    Results are given in units of fm. 
    ``Raw'' and ``GSO'' stand for results obtained before and after the Gram-Schmidt orthonormalization.
    \label{tab:charmonium_rms}
      }
      \begin{ruledtabular}                                                                              
      \begin{tabular}{lcccc} 
       State  & $\eta_c(1S)$ &$J/\psi$ & $\eta_c(2S)$ & $\psi(2S)$\\
       \hline
       $r_{\rm rms}$ [fm] (Raw)&  0.3348(2) & 0.3885(6) & 0.563(14)  & 0.612(18) \\
       $r_{\rm rms}$ [fm] (GSO)&  $\cdot\cdot\cdot$ & $\cdot\cdot\cdot$ & 0.606(4)  & 0.636(7) \\
    \end{tabular}
    \end{ruledtabular}                                                                                
\end{table}

As described in Sec.~\ref{sec2:kinetic_mass_potential}, the BS wave function for mesons
can provide more profound information about the internal structure of the quark-antiquark bound states.
We first discuss the quark kinetic mass, which can be read off from the difference of ``quantum kinetic energies"
${\bm \nabla}^2\phi_{\Gamma}/\phi_{\Gamma}$ between the members of hyperfine multiplets, as was shown in Eq.~(\ref{eq:kinetic_mass}). 
The time average for ${\bm \nabla}^2\phi_{\Gamma}/\phi_{\Gamma}$ 
appearing in Eq.~(\ref{eq:kinetic_mass}) was performed in the range 
of $[t_{\rm min}/a:t_{\rm max}/a]=[24:33]$ for the $1S$ states and [7:12] for the $2S$ states.

\begin{figure}[ht]
  \centering
  \includegraphics*[width=.50\textwidth]{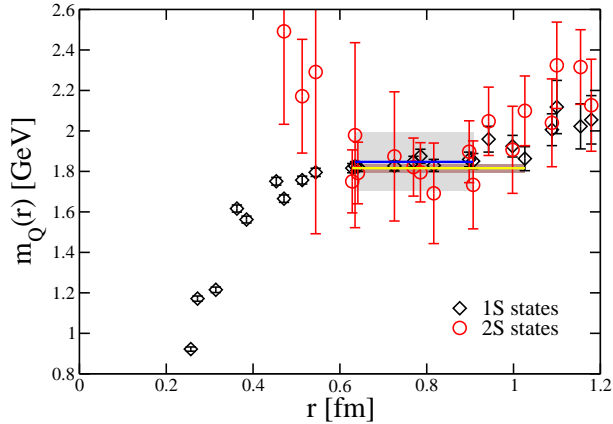} 
     \caption{The determination of quark kinetic mass within the BS amplitude method. 
     Horizontal solid lines indicate a value of quark kinetic mass obtained 
     by fitting an asymptotic constant obtained from either $1S$ or $2S$ states.
     Shaded bands indicate the fitting range and a statistical error estimated by the jackknife method.
     }
  \label{fig:quark_mass}
  \end{figure}

Figure~\ref{fig:quark_mass} shows 
that asymptotic constants obtained from the right-hand side of Eq.~(\ref{eq:kinetic_mass})
for both $1S$ and $2S$ states appear to be
overlapped in the range of $0.6~\text{fm}\alt r \alt 1.0~\text{fm}$.
A value of the kinetic mass of the charm quark is determined by a constant fit over above the $r$-range
with $\chi^2/\text{d.o.f.} < 2$.

We then obtain $\mQ(1S)=1.816(21)$~GeV from the $1S$ wave functions and $\mQ(2S)=1.847(145)$~GeV 
from the $2S$ wave functions. Both values are consistent with each other, and also with 
our previous work as listed in Table~\ref{tab:charmquark_mass}.
This indicates that within the current precision, a unique result for the quark kinetic mass 
is likely given regardless of the choice of either the ground- or excited-state pairs.
This observation is highly consistent with the success of a potential description of the charmonium system.

  %
  %

We are now ready to consider the final question of whether or not a series of the BS wave functions from the ground state to excited states is generated by the same ``nonlocal" potential. 
To answer this question is meant to justify the time-dependent BS amplitude method even for
the $\QQbar$ system. 

In order to perform rigorous comparison, 
we will determine the central spin-independent part of the interquark potential $V_C(r)$ 
from the BS wave functions of excited states in a manner independent from those of ground states.
Indeed, the quark kinetic masses have been already evaluated for both $1S$ and $2S$ states.
Therefore, the central potential $V_C(r)$ and the quark mass $\mQ$ 
can be self-consistently determined within
a single set of ${\bm r}$-dependent two-point correlation functions for each $nS$ state,
as shown in Eq.~(\ref{Eq:Center}).

%
%
 \begin{table}[ht]
  \centering
   \caption{Summary of 
   charm quark masses, which are determined from the BS amplitudes
   of both $1S$ and $2S$ charmonium states. Their fit ranges $[r_{\rm min}/a:r_{\rm max}/a]$ 
   are summarized in units of GeV.
     \label{tab:charmquark_mass}
   }
   \begin{ruledtabular}                                                         
     \begin{tabular}{cccc} 
     Method & Previous work~\cite{Kawanai:2015tga}  & \multicolumn{2}{c}{Variational method} \\
     Type of source & Wall source & \multicolumn{2}{c}{Gauss-smeared sources} \\
     State & $1S$  &  $1S$  & $2S$ \\
     \hline
    $\mQ$ [GeV]& 1.784(23) & 1.816(21) & 1.847(145)   \\
    Fit range & $[6:7\sqrt{3}]$ & $[4\sqrt{3}:8\sqrt{2}]$& $[7:10]$ \\
     \end{tabular}
   \end{ruledtabular}                                                           
 \end{table}

\begin{figure}[ht]
  \centering
  \includegraphics*[width=.50\textwidth]{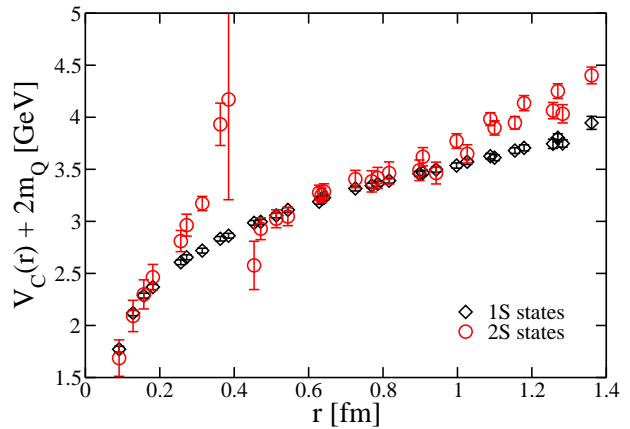} 
     \caption{Central (spin-independent) charmonium potentials calculated from
     the BS wave functions using the $S$-wave ground ($1S$) states and their first radially 
     excited ($2S$) states. For clarity of the figure, the ``threshold energy value" $2\mQ$,
     that was encoded in the constant energy shift ($E^{nS}_{\rm ave}=M^{nS}_{\rm ave}-2\mQ$),
     is not subtracted. Note that there is no adjustment parameter. 
     }
  \label{fig:central_potentials}
  \end{figure}

Figure~\ref{fig:central_potentials} shows two independent results of the central 
potential $V_C(r)$ using the BS wave functions of either $1S$ or $2S$ states. 
For clarity of the figure, the ``threshold energy value" $2\mQ$,
which is a part of the constant energy shift ($E_{\rm ave}=M_{\rm ave}-2\mQ$),
is not subtracted. The spin-averaged masses, $M_{\rm ave}(1S)$ and 
$M_{\rm ave}(2S)$, have been obtained by 
the variational method as described previously.
Thus, it should be emphasized that no adjustment constant is added for comparison. 

The gross features of the resulting central potential $V_C^{2S}(r)$ from the $2S$ states
are basically analogous to those of the $1S$ states $V_C^{1S}(r)$. 
Although data points in the intermediate ($0.5 \alt r \alt 1.1$~fm) and short-range ($r\alt 0.3$~fm) parts
of the $V_C^{2S}(r)$ agree well with a shape of $V_C^{1S}(r)$,
some discrepancy beyond the quoted statistical errors appears 
in two specific regions: around $r=0.4$~fm and at long distances ($r \agt 1.1$~fm).

The origin of the former discrepancy can be attributed to the presence of a node 
in the $2S$ wave function, which is located at $r\approx 0.4$~fm, as shown 
in Fig.~\ref{fig:BSwavefunc}. One should be reminded that the potential
defined in the BS amplitude method is basically calculated 
by the second spatial derivative of the BS wave function divided 
by the BS wave function, ${\bm \nabla}^2\phi_{\Gamma}/\phi_{\Gamma}$.
Therefore, the potential cannot be given only at nodes of the BS wave function. 

In this sense, the statistical uncertainties may lead to divergent behavior near the nodes. 
For the $2S$ wave functions, the resulting potential is rendered positively (negatively) divergent 
on the left (right) side of its singularity. This accounts for a discontinuity behavior 
appearing in $V_C^{2S}(r)$. 
Another consequence of the presence of the nodes may enhance 
a chance of unwanted excited-state contamination, since the strength
of other state contributions in ${\bm r}$-dependent two-point correlation functions 
may exceed that of the target state at its nodes.

As for another discrepancy 
found at long distances, it should be simply because of
the larger statistical uncertainties in the BS wave function of the higher-lying excited states.
As shown in Fig.~\ref{fig:BSwavefunc}, the BS wave functions of 
both $1S$ and $2S$ states are localized around the origin and vanish at long distances. 
The signal-to-noise ratio on the quantity of ${\bm \nabla}^2\phi_{\Gamma}/\phi_{\Gamma}$
becomes worse rapidly as the spatial distance $r$ increases because of the localized
nature of the BS wave functions. This tends to cause large systematic uncertainties 
at long distances, stemming from the choice of time window for the averaging process of 
${\bm \nabla}^2\phi_{\Gamma}/\phi_{\Gamma}$ over the time-slice range
(for details on the ``time-average" procedure, see Ref.~\cite{Kawanai:2013aca}). 
The time average for ${\bm \nabla}^2\phi_{\Gamma}/\phi_{\Gamma}$
was performed in the range of $[t_{\rm min}/a:t_{\rm max}/a]=[24:33]$
for the $1S$ states and [7:12] for the $2S$ states.

Indeed, the {\it string-breaking-like} behavior of the charmonium potentials found in
our previous study~\cite{Kawanai:2015tga} has, as expected, gone away 
in $V_C^{1S}(r)$, whose statistical uncertainties at long distances are much under control
in this study, due to effectively higher statistics using the average of multiple sources.  
We then conclude that the discrepancies of $V_C^{1S}(r)$ and $V_C^{2S}(r)$ appearing 
in two regions are highly associated with statistical issues on the quantity of 
${\bm \nabla}^2\phi_{\Gamma}/\phi_{\Gamma}$, particularly for the $2S$ states.
In other words, the standard errors of $V_C^{2S}(r)$ displayed in Fig.~\ref{fig:central_potentials} 
tend to be underestimated
in those regions, where the systematic uncertainties should be seriously taken into account.

To settle the above issues, we decide to utilize the time-dependent method only 
for the analysis of $V_C^{2S}(r)$, since it enables us to use the data of the $2S$ states 
in the earlier time range, where the statistical uncertainties are relatively 
under control.
It is then expected to suppress the signal-to-noise ratio on $V_C^{2S}(r)$,
and also to reduce the hidden systematic uncertainties stemming from 
a slight contamination of the $1S$ state during the averaging process 
over the time-slice range as we discussed before. 
Here, we note that this limited usage of the time-dependent method
{\it does not assume} that the BS wave functions of the $1S$ and $2S$ states 
are generated by the same potential. We rather assume that 
the third energy levels disentangled by the variational method in this study 
are associated with the $3S$ states. In addition, the nonlocal potential that 
generates the $3S$ states is identical to that of the $2S$ states.

  %
  %
  %

  %
  %
  \begin{figure}[ht]
  \centering
  \includegraphics*[width=.50\textwidth]{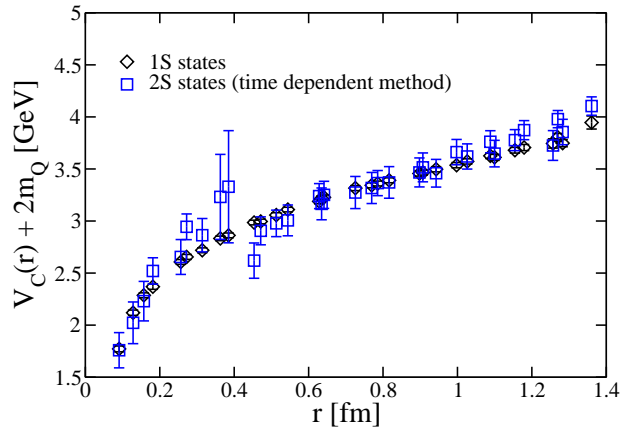} 
     \caption{Central (spin-independent) charmonium potentials from the time-independent method
     for $1S$ states and the time-dependent method for $2S$ states.
     Note that there is no adjustment parameter, the same as in Fig.~\ref{fig:central_potentials}.
     }
  \label{fig:central_potentials_with_tdep}
  \end{figure}

In Fig.~\ref{fig:central_potentials_with_tdep}, we show a comparison 
between $V_C^{1S}(r)$ from the time-independent method and $V_C^{2S}(r)$
from the time-dependent method. It again should be emphasized that no 
adjustment constant is added for comparison. 
In the determination of $V_C^{2S}(r)$, 
a change from the time-independent method to the time-dependent method
allows us to use the data in the earlier time range. 
As a result, the time-average procedure was performed in the range of $[t_{\rm min}/a:t_{\rm max}/a]=[5:11]$, which contains data points of the correlation $C_{\alpha}({\bm r},t)$ at $t/a=4$ nearest to 
the reference time ($t_0/a=3$). Recall that there are the derivative terms of $t$ in the time-dependent method.

The new result of $V_C^{2S}(r)$
using the time-dependent method fairly agrees with $V_C^{1S}(r)$.
Although a remnant of the discontinuity behavior near the node of the BS wave functions 
of the $2S$ states remains visible, the resulting charmonium potential $V_C^{2S}(r)$
exhibits linearly rising potential at large distances and Coulomb-like potential
at short distances, and is identical to $V_C^{1S}(r)$ within the current statistical precision.

  %
  %
  \begin{figure}
  \centering
  \includegraphics*[width=.50\textwidth]{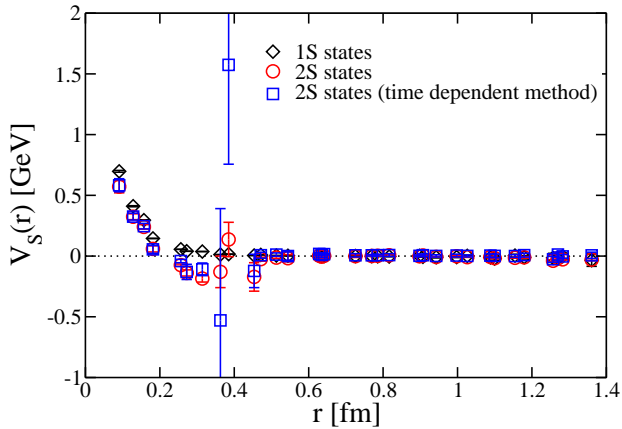} 
     \caption{Spin-spin charmonium potentials from the time-independent method 
     for both $1S$ states (diamonds) and $2S$ states (circles), and also from the time-dependent method
     for $2S$ states (squares).
     }
  \label{spin_spin_potentials}
  \end{figure}

Finally, we also determine the spin-spin potential from the BS wave functions of both $1S$ and $2S$ states.
In Fig.~\ref{spin_spin_potentials}, we compile three results of the spin-spin charmonium potential.
Open diamond symbols represent results of the spin-spin potential from the $1S$ states, $V_S^{1S}(r)$,
while results from both the time-independent (circles) and time-dependent (squares) 
methods are displayed for the spin-spin potential from the $2S$ states, $V_S^{2S}(r)$.

The potential $V_S^{1S}(r)$ exhibits a repulsive interaction for spin-triplet states
and an attractive interaction for spin-singlet states with a finite range of $r\alt 0.6$ fm,
which is the same as that discussed in Refs~\cite{{Kawanai:2011jt},{Kawanai:2015tga}}. 
On the other hand, the circle symbols of the potential $V_S^{2S}(r)$, which are given by 
the original time-independent method, reveal a small negative dip 
in the region $0.3 \alt r \alt 0.4$~fm, though the positive part of $V_S^{2S}(r)$ 
at short distances should be the dominant contribution of the finite-range spin-spin interaction.

The presence of the small negative dip makes a small difference between $V_S^{1S}(r)$ and 
$V_S^{2S}(r)$ within the time-independent approach.
However, the dip location is apparently near the node of the $2S$ wave functions.
As described previously, there is a subtlety in the calculation of 
${\bm \nabla}^2\phi_{\Gamma}/\phi_{\Gamma}$ near 
the zero of $\phi_{\Gamma}$. 
Thus, in the case of the spin-independent central potential, 
it is found that
an application of the time-dependent method is certainly effective 
in the analysis of the $2S$ states. 

Although there was no drastic change from the time-independent method to 
the time-dependent method in the case of the spin-spin potential, 
the latter result slightly becomes in agreement 
with $V_S^{1S}(r)$ within a few standard deviations. 
We may conclude that the difference between $V_S^{1S}(r)$ and 
$V_S^{2S}(r)$ is not statistically significant. We do not, however, rule out the possibility 
that different sizes of the $S$-$D$ mixing effect on the $J/\psi$ and $\psi(2S)$ states may 
lead to some difference in the spin-spin potential.

Indeed, the present calculation does not take into account the presence 
of the tensor interaction in the spin-dependent potentials, which causes
possible partial-wave mixings except for the PS channel. 
No significant difference found in both the central and spin-spin 
potentials calculated from the $1S$ and $2S$ states suggests that the possible $S$-$D$ 
mixing is not a leading effect for both the $J/\psi$ and $\psi(2S)$ states.

\section{Summary} 
We have calculated the BS wave functions for both the ground and
first excited states of the $S$-wave charmonia ($\eta_c$ and $\psi$ mesons)
in full lattice QCD. Our simulations have been carried out with the RHQ action for 
the charm quark ($M_{\eta_c}\approx 2985$ MeV and $M_{J/\psi}\approx 3099$ MeV) 
on the (2+1)-flavor PACS-CS gauge configurations near the physical 
point ($M_{\pi}\approx 156$ MeV).

The optimal charmonium operators have been successfully obtained 
for the ground and first excited states of the $S$-wave 
charmonia, using the variational method by means of a set of 
basis meson operators that are composed of spatially smeared quark sources 
with four successive smearing radii. We then calculated the BS wave functions of both the
$1S$ and $2S$ charmonium states. 
Compared with the results of $1S$ states, the BS wave functions of both the $\eta_c(2S)$ 
and $\psi(2S)$ states exhibit a specific nodal structure in the radial direction.

Although the orthonormality condition is slightly violated between 
the resulting $1S$ and $2S$ wave functions, there is only a slight difference 
in the profile shapes between the $2S$ wave functions obtained 
before and after the Gram-Schmidt orthonormalization (GSO). 
In either case, 
it is observed that the $2S$ wave functions $\phi^{2S}_{\Gamma}(r)$ are slightly extended in space 
in comparison to the $1S$ wave functions $\phi^{1S}_{\Gamma}(r)$. Indeed, we obtain a relatively larger 
spin-averaged value of $(r_{\rm rms})_{\overline{2S}}\sim 0.60$~fm (before GSO) and 
$0.63$~fm (after GSO) for the $2S$ states 
in comparison to that of the $1S$ states, $(r_{\rm rms})_{\overline{1S}}\sim 0.38$~fm.

We have read off the value of the charm quark mass from the long-distance asymptotic 
value of the difference of ``quantum kinetic energies," ${\bm \nabla}^2\phi_{\Gamma}/\phi_{\Gamma}$, 
between the members of hyperfine multiplets.
It is found that the resulting charm mass is consistent regardless of the choice of 
either the ground- or excited-state pairs in the $S$-wave charmonia.

Both the spin-independent central [$V_C^{2S}(r)$] and spin-spin [$V_S^{2S}(r)$] parts 
of the interquark potential determined from $\phi^{2S}_{\Gamma}(r)$ within 
the time-independent BS amplitude method are basically analogous to those 
of the $1S$ states, $V_C^{1S}(r)$ and $V_S^{1S}(r)$.
The large discrepancies are limited in the particular region,
where subtlety is involved in the calculation of ${\bm \nabla}^2\phi_{\Gamma}/\phi_{\Gamma}$
due to the almost zero value of $\phi_{\Gamma}$ that happens near the node of $\phi^{2S}_{\Gamma}(r)$
or at long distances with large statistical uncertainties. 

To overcome the statistical issues on $\phi^{2S}_{\Gamma}(r)$, the new time-dependent 
BS amplitude method was applied only for the analysis of both $V_C^{2S}(r)$ and $V_S^{2S}(r)$.
The spin-independent central potential $V_C^{2S}(r)$ is identical to $V_C^{1S}(r)$ within the current statistical precision, while the discrepancy between the spin-spin potentials, $V_S^{1S}(r)$ and $V_S^{2S}(r)$, 
still remains more or less visible near the node location.

Although we do not rule out the possibility that different sizes of the $S$-$D$ 
mixing effect on the $J/\psi$ and $\psi(2S)$ 
states may lead to some difference in the spin-spin potential, the difference 
between $V_S^{1S}(r)$ and $V_S^{2S}(r)$ is not statistically significant. 
Therefore, our results suggest that the possible $S$-$D$ mixing,
which is assumed to be negligible in our current analysis, is not a leading effect for both 
the $J/\psi$ and $\psi(2S)$ states. 

We thus conclude that a universal interquark potential and a unique quark mass
can be simultaneously defined in a series of the BS amplitudes from the ground state
to excited states. What this means is two-fold: (1) it ensures the reliability of the time-dependent approach 
in the BS amplitude method for the $\QQbar$ system, and (2) it strongly supports the validity of the potential description 
for the charmonium system, at least below the open-charm threshold.

We plan to extend our research to determine all spin-dependent potentials including
the tensor and spin-orbit forces and also intend to take into account the $S$-$D$ 
mixing effect on the $J/\psi$ and $\psi(2S)$ states in future analysis.


\begin{acknowledgments}
We thank the PACS-CS Collaboration for making 
their (2+1)-flavor gauge configurations available through ILDG/JLDG.
Numerical calculations reported here 
were partially carried out on the COMA (PACS-IX) system at the CCS, 
University of Tsukuba, and LX406Re-2 at the Cyberscience Center, Tohoku University.
\end{acknowledgments}


\end{document}